\documentclass[preprint2]{aastex}
\usepackage{CJK}

\shorttitle{Polarization Sequence for SNe Ia} \shortauthors{Meng,
Zhang \& Han}

\begin{document}


\title{A Polarization Sequence for Type Ia Supernovae?}


\author{X. Meng £¨ÃÏÏé´æ£©$^{\rm 1,2,3}$, J. Zhang £¨ÕžӼף©$^{\rm 1,2,3}$, Z. Han £¨º«Õ¼ÎÄ£©$^{\rm 1,2,3}$}
\affil{$^{\rm 1}$Yunnan Observatories, Chinese Academy of Sciences, 650216 Kunming, PR China\\
$^{\rm 2}$ Key Laboratory for the Structure and Evolution of
Celestial Objects, Chinese Academy of Sciences, 650216 Kunming, PR
China\\
$^{\rm 3}$Center for Astronomical Mega-Science, Chinese Academy of
Sciences, 20A Datun Road, Chaoyang District, Beijing, 100012, P.
R. China} \email{xiangcunmeng@ynao.ac.cn}





\begin{abstract}
Early polarization observations on Type Ia supernovae (SNe Ia) may
reveal the geometry of supernova ejecta, and then put constraints
on their explosion mechanism and their progenitor model. We
performed a literature search of SNe Ia with polarization
measurements and determined the polarization and relative
equivalent width (REW) of Si II 635.5-nm absorption feature at -5
days after the maximum light. We found that either the
distribution of observed polarization degree is bimodal, i.e. the
broad line SNe Ia have systematically higher polarization than all
other SNe Ia, or all kind of SNe Ia share the same polarization
sequence, i.e. the polarization of Si II 635.5-nm absorption
feature increases with the REW. We also discussed the potential
meaning of the discovery on the explosion mechanism and progenitor
model of SNe Ia.

\end{abstract}





\section{INTRODUCTION}
\label{sect:1}There is a consensus that type Ia supernovae (SNe
Ia) are thermonuclear explosions of carbon-oxygen white dwarfs (CO
WDs) in binary systems (\citealt{WANGB12}; \citealt{MAOZ14}).
Although they are used empirically to measure the cosmological
parameters, which resulted in the discovery of the accelerating
expansion of the universe (\citealt{RIE98}; \citealt{PER99}),
their progenitor nature and explosion mechanism are still unclear
(\citealt{HN00}; \citealt{LEI00}; \citealt{MENGXC15}).

A CO WD in a binary system accretes material from its companion to
increase its mass to a maximum stable mass, where an explosive
nuclear burning is ignited, and then the WD explodes as a SN Ia,
leaving no remnant (\citealt{HN00}; \cite{WANGB12};
\citealt{MAOZ14}). There is a decade-long debate on the companion
nature, and on how the explosive nuclear burning is triggered and
how burning front propagates through the WD. The resulting
chemical structures are dramatically different from various
models, which may leave some essential information via a polarized
spectrum (\citealt{WANGLF08}). Models successfully reproducing the
spectrum of SNe Ia generally start from a subsonic nuclear
burning, or deflagration, but there are endless arguments on
whether the burning front becomes supersonic after the early
deflagration, and recently on where there is a deflagration
(\citealt{KHOKHLOV91}; \citealt{NTY84}). Generally, a long
deflagration phase may result in more chemical clumps within a
wider velocity range (\citealt{REINECKE02}; \citealt{BLONDIN11}).
Spectropolarimetry and spectral analysis may probe clump structure
and velocity distribution of supernova ejecta, which may constrain
the explosion model, even the progenitor model of SNe Ia, although
all extragalactic supernovae remain point-like in the sky
(\citealt{WANGLF08}).

It is well known that SNe Ia show spectroscopic, photometric and
spectropolarimetric diversity (\citealt{FIL97}; \citealt{LEI00};
\citealt{WANGLF07}; \citealt{BRANCH09}). Among normal SNe Ia,
these properties are correlated to form 1 parameter sequences,
although the sequences may not be perfectly correlated. Various
peculiar SNe Ia sub-classes buck some, but not all, of these
trends. For example, it has been shown that the polarization of Si
II 635.5-nm absorption feature at -5 days after the maximum light
correlates with the light curve width parameter
(\citealt{WANGLF07}), and also that the polarization correlates
with the velocity-gradient parameter, $\dot{v}_{\rm Si~II}$,
inferred from the Si II 635.5 nm absorption feature
(\citealt{MAUND10a}). However, there are always some SNe Ia
deviating from these correlations. It becomes significantly
meaningful that whether or not there exists a sequence, in which
all kind of SNe Ia follow the sequence. We try to find such a
sequence.

In section \ref{sect:2}, we simply describe our method, and
present the results in section \ref{sect:3}. In section
\ref{sect:4}, we show discussions and conclusions.

\begin{table*}[htbp]
\caption[]{The information of SNe Ia shown in Fig.~\ref{prw6new}.
The second and the fourth rows are the phases of polarization and
spectrum observations. $P_{\rm Si}$ is the polarization degree of
Si II 635.5 nm absorption line, pEW is the pseudo-equivalent width
and $a$ is the relative depth of the same line. The seventh row
shows the Branch classification of the SNe Ia and the last row is
the references. In the last row, LJT means that the spectrum is
obtained from the LiJiang 2.4-m telescope at Yunnan Observatories.
S-Ch means super-Chandrasekhar.}\label{Tab:1}
\begin{center}
\begin{tabular}[b]{cccccccc}  
\\
\hline
SN & Phase            & $P_{\rm Si}$ & Phase         & pEW($\mathring{\rm A}$)    &$a$& Branch &Ref \\
   & for polarization &              & for spectrum  & of 6355 $\mathring{\rm A}$ &   & type   &  \\
\hline
1996X  & -4.2 & 0.50(20) & -4  & 83.22 & 0.61 & CN & 1, 2 \\
1997bp & -5.0 & 0.90(10) & -3  & 184.1 & 0.74 & BL & 1, 2 \\
1997bq & -3.0 & 0.40(20) & -6  & 175.5 & 0.71 & BL & 1, 2 \\
1997br & -2.0 & 0.20(20) & -7  & 18.60 & 0.14 & SS & 1, 3 \\
1999by & -2.5 & 0.40(10) & -4  & 87.09 & 0.58 & CL & 1, 2 \\
2001V  & -7.3 & 0.00(07) & -5  & 56.08 & 0.35 & SS & 1, 2 \\
2001el & -4.2 & 0.45(02) & -4  & 93.70 & 0.51 & CN & 1, 4 \\
2002bo & -5.0 & 0.90(05) & -5  & 156.10& 0.71 & BL & 1, 2\\
2002el & -6.4 & 0.72(09) & +12 & 123.8 & 0.68 & CL & 1, 3 \\
2002fk & -5.5 & 0.67(10) & -3  & 64.83 & 0.49 & CN & 1, 2 \\
2003W  & -4.5 & 0.64(10) & -5  & 152.30& 0.59 & BL & 1, 3 \\
2004dt & -7.3 & 1.60(10) &-6.5 & 200.80& 0.71 & BL & 1, 3 \\
2004ef & -4.1 & 1.10(30) & -4  & 132.90& 0.61 & BL & 1, 3 \\
2004eo & -5.9 & 0.71(08) &-5.6 & 103.20& 0.59 & CL & 1, 3 \\
2005cf & -9.9 & 0.44(05) & -5  & 88.80 & 0.50 & CN & 1, 5 \\
2005de & -4.4 & 0.67(14) & -1  & 102.80& 0.67 & CL & 1, 3 \\
2005hk & -4   & 0.36(17) &-4.3 & 20.69 & 0.16 & SS & 3, 6 \\
2005ke & -7   & 0.39(08) &  0  & 102.1 & 0.62 & CL & 2, 7 \\
2006X  & -5   & 1.00(10) & -7  & 189.4 & 0.72 & BL & 8, 9 \\
2007le & -5   & 0.85(10) & -6  & 122.2 & 0.55 & BL & 3, 10 \\
2011fe & -6   & 0.33(03) & -6  & 87.71 & 0.56 & CN & 11, LJT \\
2012fr & -5   & 0.30(05) & -4  & 70.63 & 0.45 & SS & 12, 13, LJT \\
2014J  & -3   & 0.50(10) & -3  & 111.7 & 0.62 & CN & 14, LJT \\
2016coj& -9.1 & 0.90(10) & -7  & 128.4 & 0.67 & CN & 15\\
2002bf & +3   & 0.40(10) & +3  & 170.9 & 0.77 & BL & 3, 16 \\
       &      &          & +2  & 150.8 & 0.65 &    & 3    \\
2003hv & +5   & 0.25(05) & +5  & 112.8 & 0.64 & CN & 10, 17  \\
       &      &          & +1  & 112.0 & 0.66 &    & 17  \\
2004S  & +9   & 0.26(04) & +8.3& 62.2  & 0.51 & CN & 3, 18 \\
       &      &          & +2  & 99.56 & 0.53 &    & 3 \\
2009dc & +5.6 & 0.50(10) & +7  & 58.08 & 0.44 &S-Ch& 19, 20\\
       &      &          & -4  & 59.08 & 0.38 &    & 19\\
\hline
\end{tabular}
\end{center}
\textbf{Reference:} 1. \cite{WANGLF07}; 2. \cite{BLONDIN12}; 3.
\cite{SILVERMAN12}; 4. \cite{WANGLF03}; 5. \cite{WANGXF09}; 6.
\cite{MAUND10b}; 7. \cite{PATAT12}; 8. \cite{PATAT09}; 9.
\cite{YAMANAKA09}; 10. \cite{MAUND10a}; 11. \cite{SMITH11}; 12.
\cite{MAUND13}; 13. \cite{ZHANGJJ14}; 14. \cite{PATAT14}; 15.
\cite{ZHENGWK16}; 16. \cite{LEONARD05}; 17. \cite{LELOUDAS09}; 18.
\cite{CHORNOCK08}; 19. \cite{TAUBENBERGER11}; 20. \cite{TANAKA10}.
\end{table*}

\begin{figure}
\centerline{\includegraphics[angle=270,scale=.35]{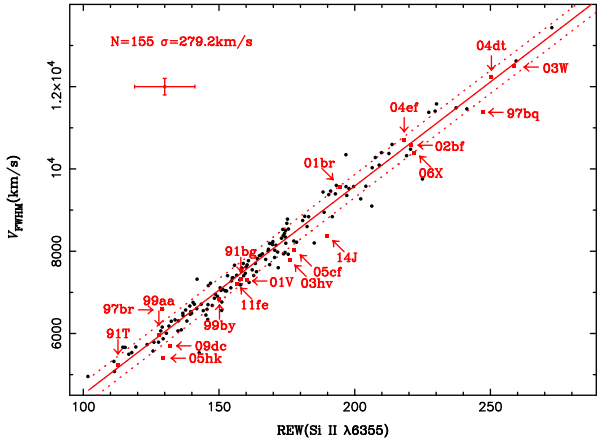}}
\caption{The correlation between the full width at half-maximum
(FWHM) intensity in velocity space and REW of Si II 635.5-nm
absorption features for 155 SNe Ia around maximum light. The red
solid line represents a linear fit, $V_{\rm
FWHM}=-536.04+50.602\times {\rm REW}$, and the dotted lines show a
statistical $1\sigma $ error of 279.2 km/s. The red cross shows
the typical measurement error of Si II absorption
feature.}\label{rrv}
\end{figure}

\section{METHOD}
\label{sect:2}

\subsection{The method to treat the data}\label{subs:2.1}
We collected the polarizational and spectral data of Si II
635.5-nm line for various SNe Ia from published literatures. All
the polarization data and most of the spectral data are collected
from the literatures and some spectrum are taken with the LiJiang
2.4-m telescope of Yunnan Observatories. The details of the data
are summarized in Tab.~\ref{Tab:1}. The sample (see
Tab.~\ref{Tab:1}) includes all sub-classes of SNe Ia, even a
peculiar object SN 2005hk (\citealt{BRANCH09}).

Generally, the peak level of the polarization of Si II 635.5-nm
line in polarization spectrum is taken as the polarization degree
of the line, and the peak value usually appears at the absorption
minimum of the line in flux spectrum. The degree of polarization
across the silicon line changes with time after explosion, and the
relation between the polarization degree and the time may be
fitted by a second order polynomial (\citealt{WANGLF07}). It is
found that the polarization degrees of some SNe Ia at -5 days
after maximum light are correlated to their maximum light
(\citealt{WANGLF07}). In this paper, for the collected
polarization data, we normalized polarization degree to the value
at -5 days after maximum light by subtracting $0.041(t + 5) +
0.013(t + 5)^{\rm 2}$ from the observed degree of polarization,
where $t$ is the day after the optical maximum light of a SN Ia
(\citealt{WANGLF07}).

In the polarization sample, the polarization observations for four
SNe Ia were carried out after their maximum light, and their
polarization degrees at -5 days after their maximum light were
estimated based on their velocity evolution of Si II 635.5-nm
absorption feature (\citealt{MAUND10a}), i.e.
\begin{equation}
P_{\rm Si~II}=0.267+0.006\times \dot{v}_{\rm Si~II},
\end{equation}
where $\dot{v}_{\rm Si~II}$ is the expansion velocity gradients of
photosphere inferred from the Si II 635.5 nm absorption feature
(see Table.~\ref{Tab:1}).

We then looked for the spectrum closest to -5 days in literatures
and defined a relative equivalent width (REW), which is defined as
the ratio of the pseudo-equivalent width (pEW) to the relative
depth ($a$) of an absorption feature. For the meaning of the REW,
please see the next subsection.  The measurements of the pEW and
$a$ for the absorption feature of Si II 635.5 nm are exactly the
same to the method in \cite{SILVERMAN12}. The REW values of the
four SNe Ia whose polarization measurements were carried out after
the maximum light are derived from the spectra closest to the time
of -5 days after their maximum light.

\subsection{The meaning of REW}\label{subs:2.2}
In this paper, REW is defined as the ratio of the
pseudo-equivalent width (pEW) to the relative depth ($a$) of an
absorption feature (\citealt{SILVERMAN12}). Actually, based on the
definition of the REW, it can always be written into a form of

\begin{equation}
{\rm REW}=c_{\rm 1}(\nu_{\rm 2}-\nu_{\rm 1})+c_{\rm 2}(v_{\rm
2}-v_{\rm 1})\label{eq:rew}
\end{equation}
in first order approximation (\citealt{RYBICKI78}), where $c_{\rm
1}$ and $c_{\rm 2}$ are coefficients which are related to the
species of an element and temperature. $\nu _{\rm 1}$ and $\nu \rm
2$ are the corresponding frequencies at the blue and red wings of
an absorption feature, and $v_{\rm 1}$ and $v_{\rm 2}$ are the
expansion velocities of inner and outer boundary of a given
element in the supernova ejecta. Generally, $v_{\rm 1}$ is equal
to the expansion velocity of photosphere. We show a simple proof
as follows. In a co-moving reference system, for a line profile
from an expanding atmosphere, the line profile can be expressed by

\begin{equation}
F(x)=\frac{1}{2}\int _{\rm 0}^{\rm R}\int _{\rm 0}^{\rm 1}I(z_{\rm
max},p,x)\mu d\mu dp,
\end{equation}
where $I$ is the specific intensity at the point ($z_{\rm max},p$)
in direction $\mu$ (\citealt{HUANGRQ98}). $z$ and $p$ are
coordinates in cylindrical coordinate and $z_{\rm max}$ is the
value at the surface of the atmosphere. $x$ is dimensionless
frequency in co-moving reference system and can be expressed as a
linear combination of frequency $\nu $ and material velocity field
$v$, i.e. $x=b_{\rm 0}+b_{\rm \nu }\nu +b_{\rm v}v$, where $b_{\rm
0}$ and $b_{\rm \nu }$ are correlated with the thermal velocity of
an species and the central frequency of a line, and $b_{\rm v}$ is
related to the thermal velocity of the species and $\mu $. For
simplicity, we normalized the continuum flux to be 1, and then the
REW is defined as

\begin{equation}
{\rm REW}=\frac{\int _{x_{\rm 1}}^{x_{\rm
2}}[1-F(x)]dx}{1-F(x_{\rm 0})}=f(x)|_{x_{\rm 1}}^{x_{\rm 2}},
\end{equation}
where $x_{\rm 0}$ is the value at linecore, and
$f(x)=\frac{\int[1-F(x)]dx}{1-F(x_{\rm 0})}$. Via a Taylor series
expansion at $x_{\rm 0}$, we may get

\begin{equation}
  \begin{array}{lc}
{\rm REW}\approx [f(x_{\rm 0})+f'(x_{\rm 0})(x-x_{\rm 0})]_{x_{\rm
1}}^{x_{\rm 2}} \\
\\
 \hspace{0.95cm}=f'(x_{\rm 0})[b_{\rm \nu }(\nu_{\rm 2}-\nu_{\rm 1})+b_{\rm
v}(v_{\rm 2}-v_{\rm 1})],
 \end{array}
\end{equation}
which shares the same form as Equation (\ref{eq:rew}). Therefore,
although the definition of REW is simple, it relates to two
physical quantities on supernova ejecta, and we expect to see a
linear relation between the REW and a kind of velocity difference
which can reflect the distribution of an element in supernova
ejecta and be derived from an absorption feature.

In Fig.~\ref{rrv}, we show the correlation between the REW and the
full width at half-maximum (FWHM) intensity of Si II 635.5-nm
absorption feature in velocity space, $V_{\rm FWHM}$, for $155$
SNe Ia around maximum light. We can see that there is a very good
linear relation between the REW and $V_{\rm FWHM}$, as expected
from the definition of the REW. Since an absorption feature may
reflect the properties of an element outside the photosphere in
the supernova ejecta of a SN Ia, REW may present the distribution
of silicon outside the photosphere in the velocity space of the
supernova ejecta.





\begin{figure}
\centerline{\includegraphics[angle=270,scale=.35]{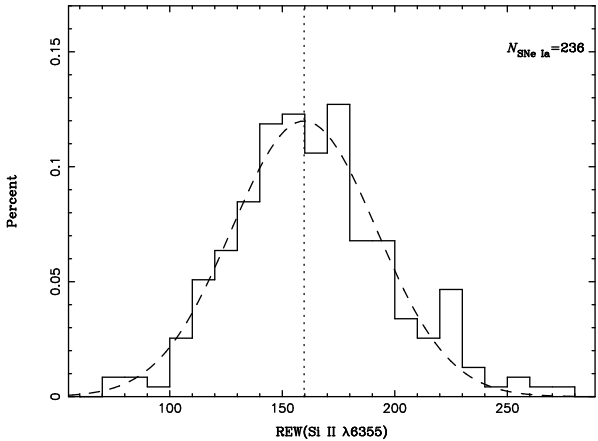}}
\caption{The distribution of REW around maximum light. The dashed
line shows the Gaussian fit of the distribution and the dotted
line shows the mean value of the REW from the fit.}\label{rrdis}
\end{figure}

\section{RESULT}\label{sect:3}
\subsection{The distribution of REW}\label{subs:3.1}
From Fig.~\ref{rrv}, we can see that the value of the REW
consecutively varies. To get the intrinsic distribution of the
REW, we collected the spectrum of SNe Ia from literatures, and the
sample is mainly from \cite{SILVERMAN12} and \cite{WANGXF13}.
Since the spectrum at -5 day after the maximum light is rare, we
just collected spectrum around maximum light. There is not any
bias when we were collecting the sample. Although it is very
likely that the value of REW may evolve with time, the
distribution of REW around maximum light may reflect the
distribution at -5 day to a great extent, at least in a sense of
the distribution's tendency and shape. In Fig.~\ref{rrdis}, we
show the distribution of the REW around maximum light. The
distribution may be well fitted by a Gaussian with an average
value of $159.7~{\rm \AA}$ and $\sigma=45.56~{\rm \AA}$, and there
is not a significant signal to show a bimodal distribution. If REW
may reflect the nature of the explosion mechanism of SNe Ia, the
one \textbf{Gaussian} distribution imply that all SNe Ia could
share the same explosion mechanism.

\subsection{The correlation between the
polarization and REW}\label{subs:3.2} 
\begin{figure}
\centerline{\includegraphics[angle=270,scale=.35]{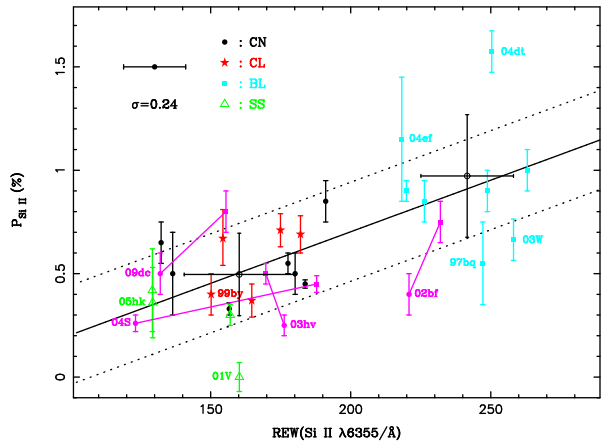}}
\caption{The correlation between the polarization and relative
equivalent width (REW) of Si II 635.5-nm line for $28$ SNe Ia.
Different color points represent different Branch types of SNe Ia,
i.e. core normal (CN, black dot), broad line (BL, blue square),
cool (CL, red star) and shallow silicon (SS, green empty triangle,
\citealt{BRANCH09}). For the SNe Ia representing by the purple
points, the polarization observations were carried our after the
maximum light (purple circle dot), and their polarization degrees
(purple square) at -5 days after the maximum light are estimated
according to their temporal velocity gradient, $\dot{v}_{\rm
Si~II}$, of Si II 635.5-nm feature (\citealt{MAUND10a}). The two
crosses show the average values for high- and low-REW groups,
respectively, and the length of the bar show the $1\sigma$ error.
The linear fit represented by the straight solid line excludes the
purple points, and the dotted lines correspond to a vertical
deviation of $1\sigma$ statistical error of $0.24\%$. The
horizontal bar at the up-left region shows the typical error of
REW, which includes the transfer errors from the measurements of
pEW and $a$.}\label{prw6new}
\end{figure}

In Fig.~\ref{prw6new}, we
show the correlation between the polarization and REW of Si II
635.5-nm line for $28$ SNe Ia. It seems that the SNe Ia are
divided into two groups based on their position in the $P_{\rm
Si}-{\rm REW}$ plane, i.e. one is a high-$P_{\rm Si}$ high-REW
group including only the broad-line (BL) SNe Ia, the other is a
low-$P_{\rm Si}$ low-REW group including core-normal (CN),
shallow-silicon (SS) and cool (CL) SNe Ia (\citealt{BRANCH09}).
This appearance is mainly derived from the gap around REW $\sim$
200 \AA. The high-$P_{\rm Si}$ high-REW group has a value of
($P_{\rm Si}$, REW) = ($0.973\%\pm0.296\%,~241.6\pm16.5$ \AA), and
the low-$P_{\rm Si}$ low-REW group has a value of ($P_{\rm Si}$,
REW) = ($0.496\%\pm0.199\%,~160.2\pm19.6$ \AA), i.e. on average,
the polarization degree of the high-$P_{\rm Si}$ high-REW group is
higher than that of the low $P_{\rm Si}$ low-REW group by about
0.5\%, and the REW value by about 80 \AA. Interestingly, the
low-$P_{\rm Si}$ low-REW group has an almost equal average value
of the REW to that of the REW distribution in Fig.~\ref{rrdis},
and the REW value of the high-$P_{\rm Si}$ high-REW group is
clearly beyond the $1\sigma$ level of the REW distribution.

The two-group appearance is mainly derived from the existence of
the gap of REW in Fig.~\ref{prw6new}. It is still not clear why
there is the gap in Fig.~\ref{prw6new}. One possible reason is
that the dependence of REW on a parameter in a successful model is
bifurcated. However, the distribution of REW is consecutive (see
Fig.~\ref{rrv}), and there is not the signal of the bifurcated
distribution of the REW. Especially, the REW distribution may well
be fitted by one Gaussian (see Fig.~\ref{rrdis}). The small size
of the sample may also contribute to the gap. To test the
probability that the gap is derived from a statistical
fluctuation, we performed a simple Monte-Carlo simulation. By
assuming a Gaussian distribution as shown in Fig.~\ref{rrdis}, we
found that the appearance of the gap in REW can be attributed to
statistical fluctuation by 19.5\% if the observational error of
the REW is not considered, while by 72.9\% if the observational
error is considered. Therefore, considering the Gaussian
distribution of the REW, the gap of REW in Fig.~\ref{prw6new} is
very likely derived from the small size of the polarization
sample. Another possible reason is from an artificial selection
effect, i.e. more attentions about polarization observations are
payed to special SNe Ia than CN SNe Ia, which could miss some CN
SNe Ia locating in the gap. For example, SN 2016coj, which is a CN
SN Ia (\citealt{ZHENGWK16}), has a REW value of $\sim~190$ \AA,
higher than any other SS, CN and CL SNe Ia in the sample here.

At the same time, there seems to be a trend that the polarization
degree increases with the REW and a linear relation may fit the
correlation, i.e.
\begin{equation}
P_{\rm Si~II}=-0.2951+4.988\times 10^{\rm -3}\times{\rm REW},
\end{equation}
where the correlation coefficient is 0.6641 and the observational
error is taken as the weight for the fitting. The vertical
deviation of $1\sigma$ statistical error for the linear fit is
$0.24\%$. For most SNe Ia, their observational error bars are
consistent with $1\sigma$ level, except SN 2004dt and SN 2001V. At
present, the sample size is small and the observational error is
still large, which affect the confident level of the correlation
between the polarization and REW as shown by the correlation
coefficient. However, combining with the Gaussian distribution of
REW, it is still very possible to exist such a correlation based
on the present data, e.g. the average value of the polarization
degree of the high-$P_{\rm Si}$ high-REW group is significantly
higher than that of the low $P_{\rm Si}$ low-REW group, although
the error bar is partly overlapped. Moreover, even the three most
deviate SNe Ia, SN 2001V, 2004dt and 2004ef, are removed, we still
can not use a horizontal line to fit the data points, but a
tighter ($\sigma=0.15\%$) and smaller slope ($3.250\times 10^{\rm
-3}$) line may fit the date points well.

In Fig.~\ref{prw6new}, the polarization observation for four SNe
Ia were carried our after the maximum light (purple points in
Fig.~\ref{prw6new}). Their positions in the $P_{\rm Si}-{\rm REW}$
plane still follow the polarization sequence except that the
polarization of SN 2009dc seems to slightly beyond the $1\sigma $
range (but still consistent with the sequence within observational
error). SN 2002bf belongs to BL sub-class, and SN 2003hv and 2004S
to CN sub-class.

\section{DISCUSSIONS AND CONCLUSIONS}\label{sect:4}
\subsection{Two polarization group?}\label{subs:4.1}
In Fig.~\ref{prw6new}, the SNe Ia seem to be divided into a
high-$P_{\rm Si}$ high-REW group and a low-$P_{\rm Si}$ low-REW
group, where the high-$P_{\rm Si}$ high-REW group are only
included the BL SNe Ia. \cite{MAUND10a} also found that among
normal SNe Ia, the high-velocity-gradient (HVG) SNe Ia show a
higher polarization than the low-velocity-gradient (LVG) SNe Ia.
Generally, the HVG SNe Ia classified by \cite{BENETTI05} overlap
with the BL SNe Ia classified by \cite{BRANCH09}, except some
peculiar SNe Ia. For example, a HVG SN Ia in Benetti's
classification, SN 2001V, belongs to SS SNe Ia in Branch's
classification, and SN 2001V has a very low polarization. Here,
the high-$P_{\rm Si}$ high-REW group consists exclusively of BL
SNe Ia. At early phase, BL SNe Ia often exhibit a high and
low-velocity components in their absorption profiles, and the
high-velocity component has the high polarization
(\citealt{MAUND13}). However, the high-velocity feature (HVF) is a
ubiquitous property of SNe Ia (\citealt{MAZZALI05a,MAZZALI05b}).
For example, SN 2012fr, which belongs to the SS subclass, shows
the clear high-velocity components of the spectral features, but
its polarization at 5 days before the maximum light is not as high
as those BL SNe Ia. Why don't the HVF in SS, CL and CN SNe Ia show
such high polarization shown in BL SNe Ia?

If the appearance of the two groups in the $P_{\rm Si}-{\rm REW}$
plane is the exact nature for SNe Ia, the results imply that SNe
Ia are derived from at least two explosion models (\citealt{HN00})
or two progenitor models (\citealt{WANGXF13}), e.g. the violent
merger model (\citealt{PAKMOR10}) produces the high-$P_{\rm Si}$
high-REW SNe Ia, while the Chandrasekhar-mass model
(\citealt{KHOKHLOV91}) and the double-detonation model
(\citealt{LIVNE90}) produce the low-$P_{\rm Si}$ low-REW SNe Ia.
However, numerical simulations show that the polarization degree
predicted from the violent model is much higher than that of the
high-$P_{\rm Si}$ high-REW SNe Ia, and the Chandrasekhar mass
model may well reproduce the polarization distribution of all SNe
Ia (\citealt{BULLA16a,BULLA16b}). As discussed in section
\ref{sect:3}, the two-group appearance is mainly derived from the
existence of the REW gap in Fig.~\ref{prw6new}, but the gap is
very likely to attribute to the small sample size.

\subsection{A polarization sequence?}\label{subs:4.1}
The polarization of silicon line reflects the asymmetry of the
silicon distribution in supernova ejecta, while the REW is a
measure of the velocity difference of silicon layer in velocity
space. Even for a spherically symmetric structure of supernova
ejecta, the different REW values are still expected from different
supernovae due to their different explosion energy. So, the
polarization degree and the REW are independent parameters, and a
correlation between the two parameter is not necessary. Therefore,
if there exists a correlation between $P_{\rm Si}$ and REW and the
correlation shown in Fig.~\ref{prw6new} is intrinsic for all SNe
Ia, the correlation will put strong constraints on the successful
explosion models of SNe Ia, since all kinds of subclasses of SNe
Ia seem to obey the same sequence, which need to be explained by
any successful model. At present stage of theoretical modelling,
deflagration (\citealt{NTY84}), delayed-detonation
(\citealt{KHOKHLOV91}, \citealt{LIVNE99}), gravitationally
confined detonation (GCD, \citealt{PLEWA04}), detonating failed
deflagration (DFD, \citealt{PLEWA07}), and violent-merger
(\citealt{PAKMOR10}) models can all plausibly be argued to produce
clumps in the ejecta, which may contribute to the polarized
absorption lines.

As shown in section \ref{subs:2.2}, the REW of an absorption line
is proportional to a kind of velocity difference which can reflect
the distribution of an element in supernova ejecta and be derived
from an absorption feature. As expected, the REW of Si II 635.5-nm
absorption feature is indeed proportional to the full width at
half-maximum (FWHM) intensity of absorption feature in velocity
space. Therefore, if the correlation between $P_{\rm Si}$ and REW
is intrinsic for all SNe Ia, we may expect that the more the
asymmetric of the distribution of silicon in the supernova, the
wider the velocity interval of silicon layer. \emph{At present, no
any simulation clearly shows such correlation in literatures}.
Whatever, some simulations showed that the delayed-detonation
model is an interesting one and may give a reasonable physical
explanation in principle. In addition, one fact must be kept in
mind that the polarization degree of the continuum of a SN Ia is
generally very low, i.e. generally less than 0.2\%, which may
exclude any progenitor model or explosion model predicting a
highly asymmetric distribution of supernova ejecta
(\citealt{WANGLF08}).

Generally, for the delayed-detonation model, a strong detonation
is more likely to produce a less turbulent silicon layer, and a
later detonation, i.e. the deflagration flame has propagated
sufficiently close to the low-density part of a expanding white
dwarf, has to bear the imprint of the complex structure of the
prior deflagration. It is then expected that the earlier the
transition from deflagration to detonation occurs, the less
asymmetric the silicon is. At the same time, the earlier the
deflagration-detonation transition occurs, the narrower the
velocity interval of silicon layer is, as shown by some numerical
simulations (\citealt{BLONDIN11,GAMEZO05}). Therefore, it is
expected that there exists a correlation between the polarization
degree of silicon lines and the velocity interval of silicon
distributed in supernova ejecta, as deduced from
Fig.~\ref{prw6new}. So, the delayed-detonation model may
qualitatively explain the correlation in Fig.~\ref{prw6new}.

\begin{figure}
\centerline{\includegraphics[angle=270,scale=.35]{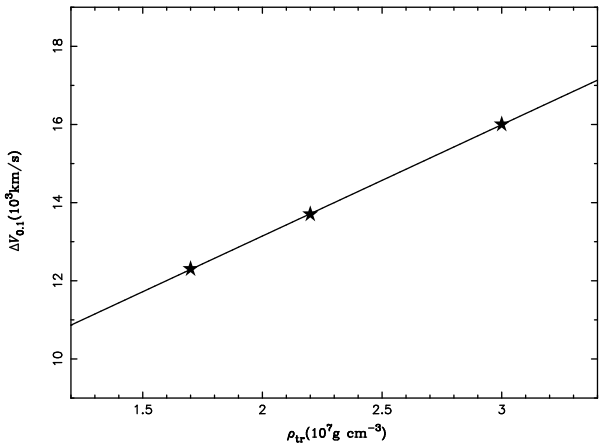}}
\caption{The correlation between the velocity interval of silicon
layer in supernova ejecta and the transition density from
deflagration to detonation, where the star points are obtained
from 1D numerical simulations (\citealt{IWAMOTO99}). The velocity
interval is defined as the velocity difference between the two
velocity point, where the mass fractions of the silicon is $0.1$.
The solid line is the linear fit of the points, $\Delta V_{\rm
0.1}=7.4477+2.8488\times\rho _{\rm tr}$, where $\Delta V_{\rm
0.1}$ is in unit $10^{\rm 3}$ km/s and $\rho _{\rm tr}$ in
$10^{\rm 7}$ {\rm g~cm$^{\rm -3}$}.}\label{theory}
\end{figure}

Different models use different free parameters to denote the
transition from deflagration to detonation. Multi-dimensional
numerical simulations generally used a transition time $t_{\rm
tr}$ from deflagration to detonation as a free parameter
(\citealt{KASEN09,GAMEZO05}), which translates into a transition
density $\rho_{\rm tr}$ from deflagration to detonation as the
free parameter in 1D simulations. The models with different
$\rho_{\rm tr}$ in 1D simulations may provide excellent fits to SN
Ia spectra and light curves and account for the energetics of SNe
Ia (\citealt{HN00}). If the delayed-detonation may explain the
correlation in Fig.~\ref{prw6new}, a correlation between the free
parameter denoting the transition from deflagration to detonation
and the velocity interval of silicon layer in supernova ejecta
could be expected. To test this idea, we search for the silicon
distribution in velocity space from published literature. The
distribution should fulfil the following two criterions: 1. it is
the silicon distribution presented in the whole velocity interval,
and 2. at least three value of the free parameter was adopted. We
found that the 1D simulation in \cite{IWAMOTO99} completely
fulfils the above criterions. Here, based on the result of a 1D
numerical simulation (\citealt{IWAMOTO99}), we define the velocity
interval of the silicon layer in supernova ejecta as the velocity
difference between the two velocity points, where the mass
fraction of the silicon is 0.1 (see Figure 25 in
\citealt{IWAMOTO99}). In Fig.~\ref{theory}, we show the
correlation between the velocity interval of silicon $\Delta
V_{\rm 0.1}$ as the function of the transition density $\rho_{\rm
tr}$. We can see from the figure that the correlation is well
fitted by a linear relation. As shown in section~\ref{subs:2.2},
REW presents the velocity interval of the silicon layer above
photosphere. At -5 day, the location of the photosphere is still
inside the silicon layer, and then the REW is not the exact
measure of the velocity interval of the silicon layer in supernova
ejecta. However, the REW may still represent the velocity interval
to a great extent since rather a part of silicon layer is on top
of photosphere at -5 day, especially for those high-velocity part
which contributes to a sizable part of the velocity interval of
silicon layer (\citealt{IWAMOTO99}; \citealt{BENETTI05}).
Therefore, although it is not the exact same, $\Delta V_{\rm 0.1}$
defined here may represent $V_{\rm FWHM}$ to a great extent.
Combining the linear relation between the REW and $V_{\rm FWHM}$,
we expect a linear relation between the REW and the $\rho_{\rm
tr}$. Considering that the free parameter of the $\rho_{\rm tr}$
in 1D simulation plays a similar role to the numerical simulation
results of explosion model as the transition time $t_{\rm tr}$
from deflagration to detonation in multi-dimensional numerical
simulations, a linear relation between the $t_{\rm tr}$ and the
REW is expected. Especially, the later the $t_{\rm tr}$, the
closer to WD surface the deflagration, and then more complex
density structure the following detonation has to bear. In
addition, the later the $t_{\rm tr}$, the wider the velocity
interval of silicon layer is (\citealt{BLONDIN11,GAMEZO05}).
Therefore, a high polarization degree and a large REW are
simultaneously expected from a later transition from deflagration
to detonation. Moreover, since it is close to a spherically
symmetric structure for the supernova ejecta from the
delayed-detonation model, the low polarization degree of the
continuum is also not a problem for delayed-detonation model.

The above discussion is a deduction based on some present
numerical simulations and an assumption that $\Delta V_{\rm 0.1}$
may represent $V_{\rm FWHM}$ to a great extent. Especially, we
also notice that the simulations on the delayed-detonation models
are not always consistent with each other. For example, the 3D
delayed-detonation model in \cite{SEITENZAHL13} shown that the
models with weaker deflagration have more asymmetric silicon
layers, which seems to be exactly opposite to the expectations
from Fig.~\ref{prw6new}. Moreover, at -5 days after the maximum
light, the Si II line widths are not exactly measures of the
velocity intervals of silicon layer in supernova ejecta. So, the
above discussion on the delayed-detonation model is just a
qualitative analysis. More quantitative simulation is needed to
check whether or not the delayed-detonation model may reproduce
the correlation between the polarization degree and REW as shown
in Fig.~\ref{prw6new}.

\subsection{The origin of the scatter}\label{subs:4.3}
However, there is still a large scatter of $\sigma =0.24\%$ for
the correlation in Fig.~\ref{prw6new}. For most SNe Ia, their
observational error bars are consistent with $1\sigma $ level,
except SN 2004dt and SN 2001V. The origin of the scatter is still
not clear. One possible origin is the effect of the viewing angle,
e.g. the level of the scatter of $\sigma =0.24\%$ for the
correlation is well consistent with that predicted by the
delayed-detonation model (\citealt{BULLA16b}). It has been
verified that the viewing angle may significantly affect the
properties of SNe Ia, such as the absolute magnitude, the time of
maximum light, the optical spectra and especially the polarization
degree of absorption features
(\citealt{KASEN04,BLONDIN11,BULLA16a}). The effect of viewing
angle on the polarization degree is not a simple monotonic
function (\citealt{KASEN04,BULLA16a}), and the level of the
polarization uncertainties due to the viewing angle is heavily
model-dependent, especially on the progenitor model. For the
single-degenerate model, even if the explosion is spherically
symmetric, supernova ejecta may still become asymmetric due to the
existence of its companion, i.e. the companion carves out a
conical hole in the supernova ejecta (\citealt{MARIETTA00};
\citealt{MENGXC07}; \citealt{GRAY16}), which leads to an
uncertainty of the polarization degree of silicon line as large as
0.5\% (\citealt{KASEN04}), consistent with the $2\sigma $ level of
the polarization sequence in Fig.~\ref{prw6new}. As a subclass of
the double degenerate model, the violent-merger model predicts
that the difference of the polarization degree of silicon line
from different viewing angels for the same SN Ia may be as large
as $~1.8$, and then the uncertainty of the polarization degree is
even higher than $3\sigma $ level obtained here
(\citealt{BULLA16a}). Considering the polarization degree of the
continuum predicted by the violent merger model is always
significantly higher than that from observations
(\citealt{BULLA16a,BULLA16b}), the violent-merger model seems not
to be a reasonable one to explain the polarization sequence here
and the low continuum polarization of SNe Ia. Similarly, the
predicted properties of SNe Ia from the WD-WD collision model is
also quite viewing-angle dependent and the supernova ejecta is
also highly asymmetric (\citealt{RASKIN09,ROSSWOG09}). Therefore,
the WD-WD collision model also has no ability to simultaneously
explain the polarization sequence and the low continuum
polarization of SNe Ia. The polarization degree of SN 2004dt and
SN 2001V is beyond $1\sigma $ range of the polarization trend, as
found in \cite{WANGLF07} and \cite{MAUND10a}. If the viewing angle
is the origin of the polarization uncertainties, the SN 2004dt and
SN 2001V would be observed along a very special viewing angle.

In addition, there are at leat two populations contributing to SNe
Ia observationally (\citealt{WANGXF13}), which suggests that at
least two progenitor scenarios contribute to SNe Ia. Since the
sequence in Fig.~\ref{prw6new} includes all subclasses of SNe Ia,
the result here implies that all SNe Ia could share the same
explosion mechanism, no matter what their progenitor models are,
if the correlation is intrinsic for all SNe Ia.

\subsection{The peculiar SN 2005hk}\label{subs:4.4}
Among the SNe Ia shown in Fig.~\ref{prw6new}, SN 2005hk was
classified as a peculiar SN Ia (\citealt{CHORNOCK06}) and was
suggested to arise from the pure deflagration model of a
Chandrasekhar-mass CO WD(\citealt{KROMER13}). Recently, the
peculiar SN Ia like SN 2005hk was suggested to arise from the
hybrid WDs where a CO core is surrounded by an oxygen-neon (ONe)
mantle (\citealt{MENGXC14,WANGB14,KROMER15}). For a low carbon
abundance in the ONe mantle, the deflagration flame may be
switched off in the mantle region and does not translate into
detonation. If the switched-off deflagration model is taken as a
special case of the delayed-detonation model, the low polarization
degree of SN 2005hk is then naturally explained. The following
detonation after a deflagration flame may play a great role on the
final density structure of supernova ejecta in velocity space,
i.e. amplifying the velocity distribution of a given element, and
then possibly the asymmetry of supernova ejecta. Due to the
absence of the amplification of the following detonation, a low
polarization degree and a low REW may be simultaneously expected
from a SN Ia arising from the switched-off deflagration model.
Whatever, a detailed numerically polarization simulation on this
suggestion is needed.

\subsection{The super-luminous SN 2009dc}\label{subs:4.5}
The polarization degree of SN 2009dc is consistent with the
$1\sigma$ range of the linear relation in Fig.~\ref{prw6new}
within observational error. SN 2009dc is a super-luminous SN Ia
and the total mass of the supernova ejecta exceeds the canonical
Chandrasekhar mass limit (\citealt{HOWELL06,YAMANAKA09b}). The
supernova was suggested to arise from the merger of a double
degenerate system, where the final exploding WD is more likely to
have a Chandrasekhar mass, surrounded by a large amount of
carbon-oxygen circumstellar medium (CSM)
(\citealt{TAUBENBERGER13}). Generally, the supernova ejecta of a
SN Ia from the double-degenerate model is highly asymmetric, and
then a high polarization degree is expected, which is inconsistent
with the low continuum polarization degree of SN 2009dc
(\citealt{TANAKA10}). Two possible reasons may contribute to the
low polarization degree of SN 2009dc. One is that its polarization
degree at -5 days after their maximum light was underestimated
since the ejecta velocity evolution of a super-Chandrasekhar-mass
SN Ia is completely different from a normal SN Ia, e.g. the ejecta
velocity of a super-luminous SNe Ia is generally much lower than
that of a normal SN Ia (\citealt{HOWELL06}). Another possible
reason is that, if the supernova arises from a Chandrasekhar-mass
WD exploding in a dense carbon-oxygen CSM, the WD could experience
a short deflagration phase, which becomes a detonation soon. Such
a short deflagration phase does not produce a highly asymmetric
distribution of silicon, either a large velocity range. At the
same time, early transition from deflagration to detonation means
a large amount of nickel-56 (\citealt{BLONDIN11}). Such a
Chandrasekhar-mass explosion is also needed by a small continuum
polarization (\citealt{TANAKA10}). If so, the high amount of
nickel-56, low polarization degree and narrow silicon absorption
feature may be explained simultaneously, i.e. even if a SN Ia
arise from a double- degenerate system with super-Chandrasekhar
mass, there is not any special thing for its explosion mechanism.

The low polarization of SN 2009dc could also arise from the effect
of viewing angle, i.e. it was observed from a special viewing
angle (\citealt{BULLA16a}). However, it seems impossible for the
effect of the viewing angle to explain the high luminosity and low
polarization degree of SN 2009dc simultaneously, i.e a high
luminosity due to the effect of viewing angle also means a high
polarization degree of silicon line
(\citealt{BLONDIN11,HILLEBRANDT07}).\\
\\
In summary, we found that either SNe Ia are divided into two
groups in $P_{\rm Si}$-REW plane, or all SNe Ia follows a
polarization sequence. However, considering the Gaussian
distribution of the REW, it is very likely that the two-group
result is derived from the small size of the polarization sample
here. If the correlation between the polarization degree and the
REW is intrinsic for all SNe Ia, it will put strong constraints on
the explosion model. At present, no any explosion model clearly
shows such a correlation, but the delayed-detonation model is an
interesting one to qualitatively explain the sequence.

\section*{Acknowledgments}
We are grateful to the anonymous referee for his/her constructive
suggestions that greatly improved this manuscript, and to Ivo
Seitenzahl for his detailed discussion on this work. This work was
supported by the NSFC (No. 11473063, 11522327, 11403096, 11390374
and 11521303), CAS light of West China Program and CAS (No.
KJZD-EW-M06-01). Z.H. thanks the support by the Science and
Technology Innovation Talent Programme of the Yunnan Province (No.
2013HA005).

\end{document}